\documentstyle[12pt,aps]{revtex}
\begin{document}
\title{Dissipative structure formation in cold-rolled Fe and Ni during heavy ion irradiation}
\author{P. Sen, G. Aggarwal and U. Tiwari }
\address{School of Physical Sciences, Jawaharlal Nehru University,
New Delhi-110067, India}
\maketitle
\vskip .5cm
\noindent
\abstract
{We report 4-probe resistivity measurements of cold-rolled Ni and Fe 
during 100 MeV oxygen ion irradiation, at 300K. The resistivity shows 
increase and saturation, marked by jumps. Employing 200 MeV silver ion 
irradiation of Fe and Si(100) and topographically identifying strain 
at an artificial interface in the latter, we assign the resistivity 
behavior to atomic rearrangements arising from dissipation of incident 
ion energy at internal interfaces of Ni and Fe, with positive feedback.}
\newpage
Dissipative processes lead to various mechanisms of energy storage
and their release in natural systems. Development of dissipative structures in
time and space are therefore a well studied
phenomenon [1-3]. Recently, formation of such structures in time have been
theoretically predicted in metals under ion irradiation~\cite{1}. Here the
irradiating ions provide a radiation damage energy, which is stored. The resulting 
periodic oscillations of temperature and defect concentration
are seen to originate from two competing processes: (1)
defect formation during irradiation and (2) defect recombination. The
latter produces heat which initiates further recombination through
enhanced defect mobility. Under favourable conditions, oscillations set in
as a result of this feedback.

According to established principles of energy transfer during ion
irradiation, defects are introduced in a material directly by displacement
of the lattice atoms and indirectly through de-excitation of the electronic
subsystem. However at high energy ($\sim$ several MeV), achieved by
accelerating heavy ions, the above regions of energy transfer can be
spatially separated. With thin specimens, processes initiated purely by
energy transfer to lattice electrons and their subsequent de-excitation
can be investigated. Such processes are called electronic energy loss or
electronic stopping (${\rm S_e}$) initiated as the ions lose energy through
excitation of lattice electrons.

It is now established that structural modifications introduced by 
electronic loss in a material range from  displacive phase transitions 
in Ti to track formation in NiZr$_2$ alloys and metallic glasses [4-6].
While lattice defects and track formation take place above a certain threshold,
phase transformations can be a sub-threshold phenomenon. Some metals like Fe, 
Co and Zr have also been found to be sensitive to ${\rm S_e}$~\cite{7}, forming point 
defects above threshold, while short range order modification is observed in
austentic alloys~\cite{8}. In general, metals showing various allotropic phases 
are seen to be prone to defect formation and change of phase. In the case
of free surfaces and interfaces, mixing takes place irrespective of the 
material's sensitivity to ${\rm S_e}$ in the bulk. Such systems reported are 
multilayers of crystalline Ni/Ti, amorphous Si/Fe and amorphous W/Si [8,9].

Thus a possibilty of dissipating electronic loss energy through atomic
displacements exists, which under favourable conditions, could lead to dissipative
structure formation. In this Letter we first show experimental evidence of features in a
4-probe resistivity measurement in Fe and Ni, under irradiation, which we
assign to formation of such structures. We show their behavior 
when scaled with specimen temperature, ion flux, sampling
current, ${\rm S_e}$ and metallic microstructure. The process leading to the
formation of dissipative structures is 
established by a novel experiment employing x-ray topographic (XRT) analysis of strain
developed across an ion irradiated interface, artificially generated in a
single crystal. The results are extended to polycrystalline Ni and Fe
which are looked upon as ``single crystals'' with internal surfaces. The
very nature of the measurement allowed us to record the entire sequence of
energy storage and release in one single experiment. To the
best of our knowledge, this is the first such observation in a metal.

Resistivity measurements were computer controlled to operate a 4-probe setup where
10~$\mu {\rm m}$ foils (15~mm long; cold rolled and 99.99\% pure from Goodfellow, UK)
 were attached to a constant flow 
cryostat and irradiated in a selected region,
5 mm wide, between the inner pair of voltage probes, while a pair of outer probes
supplied constant sampling current (I). The metallic samples are assumed to be affected predominantly 
by ${\rm S_e}$, as the irradiating ion range, R, is larger 
than the sample thickness. The relevant parameters calculated for 100 MeV 
oxygen 
irradiation of Ni (O/Ni) and Fe (O/Fe) are ${\rm R_{O/Ni}~=~33~\mu m}$ and 
${\rm R_{O/Fe}~=~37~\mu m}$ respectively. For both these systems the estimated ion
energy loss (average values for O/Ni and O/Fe are 2.25 and 2 MeV/$\mu {\rm m}$ respectively) 
due to electronic means in the 10 $\mu {\rm m}$ thick samples is
above 99\%, calculated employing standard techniques~\cite{10}.

The response voltage (RV), generated during irradiation of Fe and Ni at 300K 
with 100 MeV oxygen ions, in presence of various currents I, is 
presented in Fig.1. The irradiation time, t, is in terms of steps, each of
0.2 sec duration so that t = 0 corresponds to the time an ion beam was
switched on. The beam-off positions are aligned to t = 380 for Ni and
t = 249 for Fe. A general feature characterising the behavior of
RV with irradiation time is the following: (1) Rise in RV during
irradiation and subsequent progress in attaining a steady state, where RV
does not vary considerably and (2) when the irradiating ions are removed,
RV drops within seconds, to its pre-irradiated value. We record two
inferences at this point. Firstly, the changes in RV are typical of defect
formation during irradiation and their subsequent annihilation in absence
of the ions. And secondly, RV increases and decreases in jumps (defined as
a discontinuous rise or fall in voltage of the order of $10^{-6}$~V) whose
number has a sampling current dependence. The inset to both the figures
show these jumps. The jumps are random in nature and disappear when RV reaches 
saturation. But RV, during irradiation, does not scale with I.

We shall now establish the nature of dissipative processes operative in
the above experiment. It is known that polycrystalline materials, similar
to the ones employed for the experiment described above, contain grain
boundaries separating individual ``single crystal'' grains from each other.
The internal periodicity of an individual grain is lost at the grain
boundary. In order to study its ion irradiation behavior we artificially create
an irradiated/unirradiated interface
(called irradiation interface henceforth) in a perfect single crystal of Si
which is investigated for 
strain, employing XRT. In XRT, a reciprocal lattice vector $\vec g$ is chosen so that incident x-rays suffer
Bragg diffraction ${\vec K}_g = \vec K +\vec g $, while a lattice strain
vector $\vec b$ modifies the scattered x-ray intensity distribution. The
relative orientation of $\vec g$ and the lattice strain vector $\vec b$
provides for the intensity selection rule $\vec g. \vec b = 0$. The
crystal surface is then mapped to provide a picture of the strain
developed at the interface. The technical details have already been published~\cite{11}.

In Fig.2 we present results obtained from our XRT analysis. The irradiation interface 
is constructed by employing a masking
grid, schematically presented in Fig.2(a), with dark lines representing
the $40 \mu {\rm m}$ strips, impenetrable to the ions with 
${\rm 850\mu m \times 850 \mu m}$ transparent regions. Two sets of mutually
perpendicular strips are employed which gives the appearance of a square
mesh; the 200 MeV silver ions irradiate the mesh and the underlying Si(100) lattice
perpendicular to the plane of the paper. The strips are so oriented with
respect to the Si(100) face that one set are aligned parallel to the $\vec
g$ = (220) direction during irradiation. When the same $\vec g$ = (220)
vector is employed for XRT mapping the irradiated surface, Fig.2(b)
results. The observed parallel lines are due to strain $\vec b_1$ developed
at the irradiation interface, while the mutually perpendicular set of
lines, which were also expected, do not appear as $\vec g. \vec b_2 =0$ for
the selected vector. The proposed strain vectors $\vec b_1$ and $\vec b_2$
which explains this result are schematically drawn in Fig.2(a). 
When the same irradiated surface is selected for XRT
mapping with another vector $\vec g$ = (040), the expected crossed lines
appear (similar to the square mesh in Fig.2(a)) as $\vec g. \vec b \neq 0$ 
for the 
strain vectors $\vec b_1$ and $\vec b_2$. This is shown in Fig.2(c). It is to 
be noted here that an angle of $45^{\circ}$ is included between $\vec g$ = (040) and
$\vec g$ = (220). Also, the information presented here comes from a depth
of 7$\mu {\rm m}$ as compared to the Ag ion range of $\sim 21\mu {\rm m}$ in Si. Thus
making the above effect originate from electronic 
loss(${\rm S_e}$=12 MeV/$\mu {\rm m}$). 

The topographic results conclusively show appearance of strain
perpendicular to an irradiated interface in regions subjected to electronic
loss. Hence a stress is associated
with this strain whose origins are related to the irradiation event. As an
irradiation interface is important here, any ion/target combination would
yield similar results with varying extent of strain. We
shall apply this result to explain irradiation induced changes in the
resistivity plots (Fig.1) of Ni and Fe.

Polycrystalline materials in general and cold-rolled foils in particular
contain a large number of stationary defects. We consider an edge dislocation 
as a representative case.
In Fig.3 we show a situation where a lattice of 7 atoms continues as a
6-atom lattice, abruptly across an edge. This leaves a string of atoms,
marked A, faced with a discontinuity. These atoms will however try to 
maintain its periodicity in the lattice and move towards one of the lines
AX or AY. This will severely distort the lattice in this region.
Under irradiation, the atom string A
will be subjected to an irradiation interface and result in a stress $\vec
\tau$, marked in Fig.3. An outcome of this could be a rearrangement
of atoms near such a dislocation edge into new configurations. These
configurations would be metastable in nature and will be sensed by a
current which is driven during the rearrangement process. If stress 
$\vec\tau$ is sufficiently large, the entire string of atoms can be displaced
to collect at grain boundaries. This could modify existing internal stress,
as well as result in reorientation of grains~\cite{12}.

In order to justify that these defects, arising out of irradiation are
indeed associated with the microstructure of the material and distinct
from the well known point defects, we show scaling of RV with various
parameters in Fig.4. In Fig.4(a) we show RV vs irradiation time
plots of cold-rolled Fe
at two sample temperatures, T = 300K and T = 80K for an identical
irradiation flux ($\sim 8.036 \times 10^8 ~ions~
cm^{-2}~sec^{-1}$). $\Delta RV (= RV_{max} - RV_{t=0})$ which is a measure of the absolute amount
of defects created and $\frac{\Delta RV}{RV}$, a quantity relative to the
inherent defects at that temperature, works out to be $3.0333 \times
10^{-5} V$ and $0.035$ at 300K and 
$5.25 \times 10^{-6} V$ and $0.037$ at 80K. If RV originated due to
point defects at room temperature, both $\Delta RV$
and $\frac{\Delta RV}{RV}$ would increase at 80K, due to increase in point
defect density resulting from reduced defect mobility and hence slower
recombination. But this is certainly not observed. It should also be noted that
jumps in RV are missing at 80K. The ions employed here are 200 MeV Ag. But at 
300K, the jumps are seen to be ${\rm S_e}$ (average value $\sim 15~{\rm MeV}/\mu{\rm m}$ for Ag/Fe) 
related and ion independent (compare Figs.4(a) and 1(b)).

Fig.4(b) shows RV vs t plots for ion flux of $8.036 \times 10^8~{\rm ions~
cm^{-2}~sec^{-1}}$ and  $1.384 \times 10^9~{\rm ions~cm^{-2}~sec^{-1}}$  at room
temperature. $\frac{\Delta RV}{RV}$ almost doubles which is consistent with
larger dissipation in the latter, where the ions interact with a larger
volume of microstructure. For a direct evidence on involvement of
microstructure, RV of a cold-rolled foil is compared, under irradiation at
room temperature, before and after annealing. Fig.4(c) shows the results.
It is well known that annealing reduces stationary imperfections and a
large volume of literature exists on this topic~\cite{12}. The effect on RV during
irradiation of this annealed foil is a substantial reduction in this
measured quantity. Hence reduction in microstructural defects is directly
reflected in RV. 

Having identified and verified the contribution of microstructural defects
towards RV in the 4-probe measurement, we now proceed to analyse the
features of Fig.1 in detail. The rise in RV with irradiation is a
result of dissipation at the ``internal surfaces'' through formation of new
atomic configurations or rearrangements under irradiation, as described in
Fig.3. Similar reports of resistivity jumps have been assigned to atomic
rearrangements in the literature~\cite{13}. The arrangements we report are not permanent but open to statistical
decay. The levelling-off of RV at the jumps is then arising out of
saturation in a quantity, which is the difference
between production rate and decay rate of new configurations. These new
configurations are randomly distributed in the material, under irradiation. 
As the production rate is
constant during irradiation, it is a substantial increase in the decay
rate which triggers these jumps. Considering each rearrangement to be
equivalent to formation of one Frenkel pair (an underestimate, as several
atoms can rearrange at dislocations/grain boundries) and employing literature
estimates of resistivity increase (${\rm 3000~\mu~\Omega~cm}$) for such pair formation~\cite{14} 
the average number of decay at the jumps turns out to be $1.968 \times 10^{-2}$.
The contribution of each jump is about 6\% of the total resistivity 
increase from beam-on to saturation. Such large decay rates at any given 
point during irradiation is not  possible without positive feedback, 
where decay of some configurations induces
other configurations to decay, limited only by the availability of such
configurations.

Presence of positive feedback always opens the possibility for 
auto-oscillations which in this case would be periodic variations 
in rearrangement
concentration and temperature. During the steady state, between beam-on
and beam-off, such structures fill up the irradiation volume. On beam-off,
these decay and is marked by steps in RV. Conceptually the release of such
structures are similar to ones observed in neutron irradiated ${\rm CH_4}$~\cite{2},
where heat had to be applied to observe them as the structures were frozen
at low temperature (below 15K). In this experiment however, the structures are
populated at room temperature and can be observed by simply switching off
the irradiation source.

Finally, the above discussion of feedback assumes the microstructure to
completely revert to its original configuration. This is however an ideal
situation. In reality, complete reversal may not be achieved and
reorientation of grains can result due to reasons already stated. 
We show in Fig.5a, $\theta$-2$\theta$ x-ray diffraction (XRD) scans for Ni.
Comparing the XRD plots 
before irradiation (BF) and after irradiation (AF), it is clear that the 
higher order reflections increase in relative intensity. Also compared in the
inset to this figure are step scans of Ni for the (200)
reflection. A systematic shift in the peak positions towards 
lower angles are recorded. Peak shifts in thin Ni films have been observed 
before, accompanied by incerase in peak widths ~\cite{15}. These were 
explained in terms of grain size variation. As we observe no 
variation in peak shape, the shifts in peak position reflect increase 
in lattice constant due to changes in existing internal stress; 
incorporation of solute(oxygen) atoms are ruled out as ion ranges far 
exceed Ni and Fe foil thicknesses. Another result obtained from XRD data of 
Fig.5 is an increase in relative intensity of all peaks towards higher angles. 
This is entirely an irradiation related effect and independent of the material 
under study or their initial relative peak 
intensity. These are assigned to texturing effects ~\cite{12}.

In conclusion, we have shown evidence of dissipative structure formation
in metals under irradiation. These are formed as rearrangements in an
already present microstructure filled with stationary imperfections like
dislocations and grain boundaries. Employing topography we have identified
the process leading to these rearrangements whose resistivity behaviour is
similar to defect formation. We have also shown the presence of positive
feedback during the process of irradiation and the resultant decay of
irradiation induced structures, in one single experiment. The unmistakable
participation of the microstructure in verifying the physical principles of
positive feedback, under irradiation, has been established.

Acknowledgement: We acknowledge the cooperation extended by NSC, New Delhi, 
SSPL, Delhi and IUC, Indore during this study. Partial financial support from
UGC is also acknowledged.

\newpage
\centerline{\bf Figure Captions}
\begin{itemize}
\item Fig.1~: The response voltage (RV) generated during irradiation by
a constant flux ($5 \times 10^9~~{\rm ions~cm^{-2}~sec^{-1}}$) of 100 MeV oxygen ions in a 4-probe resistivity
measurement setup. RV is plotted as a function of irradiation time
interval (t) when the data was collected. All plots are normalised to RV
at t~=~0 for a particular current (I). (a) For Ni
foils, the ion beam was switched on at t~=~0 and switched off at t~=~380 ;
the inset shows one jump after beam off between t~=~386 and t~=~406. (b)
For Fe foils, the ion beam was switched on at t~=~0 and switched off at
t~=~249 ; the inset shows one jump just after beam on between t~=~22 and
t~=~52 
\item Fig.2~: (a) Dark lines show two parallel sets of Ni wire strips,
attached to each other after rotation by $90^{\circ}$. Each strip of
square cross-section allows 200 MeV Ag ions to pass through the
intermediate region and irradiate a Si(100) crystal below. The strips are
so oriented with respect to Si(100) that one set, marked h, are parallel
to its $\vec g$(220). The strain vectors $\vec b_1$
and $\vec b_2$ are present on the Si surface in the configuration shown
and are obtained from the XRT results. (b) XRT employing $\vec g$(220), of
Si(100) after irradiation; one set of lines disappear as $\vec g . \vec
b_2$ = 0 (c) XRT of the same employing $\vec g$(040). Both sets of lines
appear as $\vec g. \vec b \neq 0$ for $\vec b_1$ and $\vec b_2$.
\item Fig.3~: Schematic diagram of dislocation edge and the effect of ion
irradiation as derived from the XRT results (see text). The ions enter the
plane of the paper.
\item Fig.4~: Effect on RV of an Fe foil is shown under various
conditions. (a) Dependence on temperature, (b) dependence on the ion
current and (c) dependence on annealing. This data is taken with 200 MeV
Ag ions showing ion-independent behavior; I = 100 mA.
\item Fig.5~: XRD plots before (BF) and after (AF) irradiation are
compared. (a) For Ni foils showing changes in relative intensity ; inset
shows details of the (200) reflections. (b) For Fe foils showing changes
in relative intensity ; inset shows details of the (110) reflections.
\end{itemize}
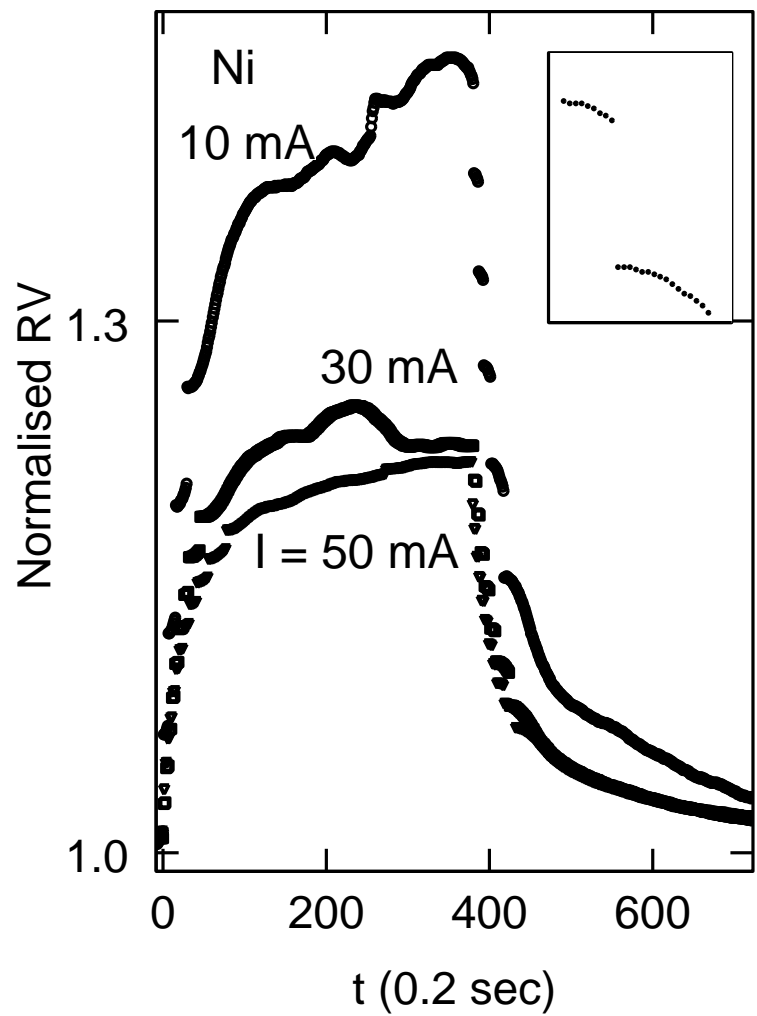
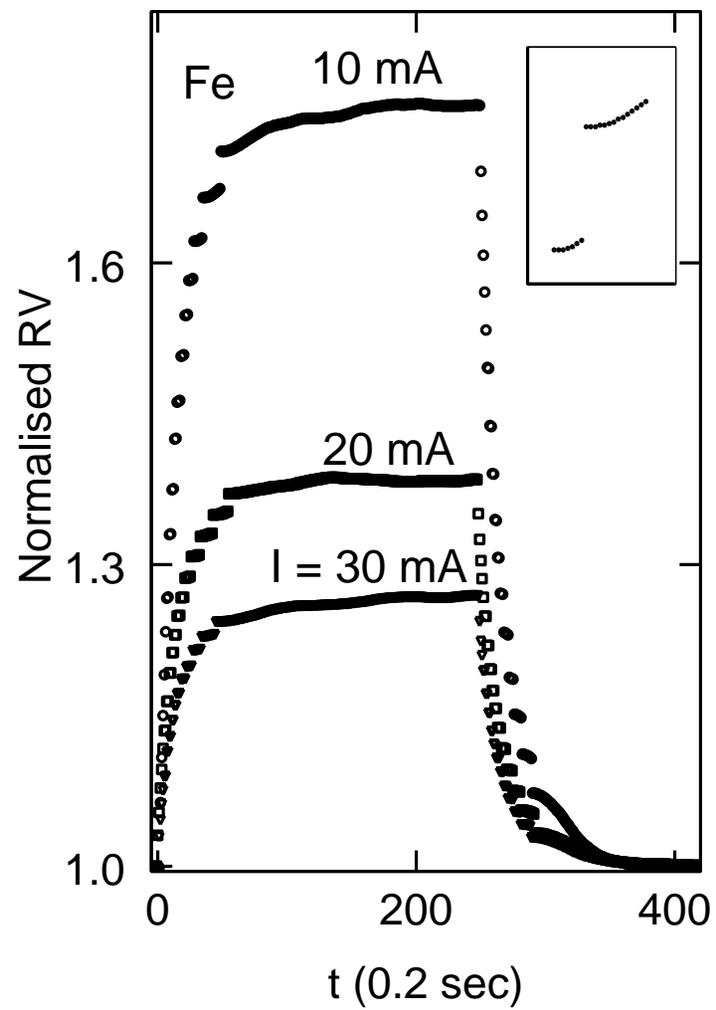
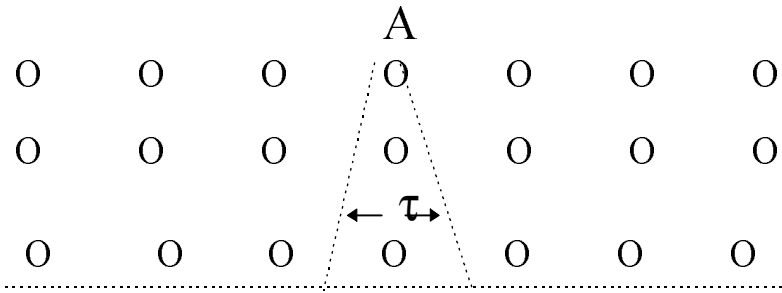
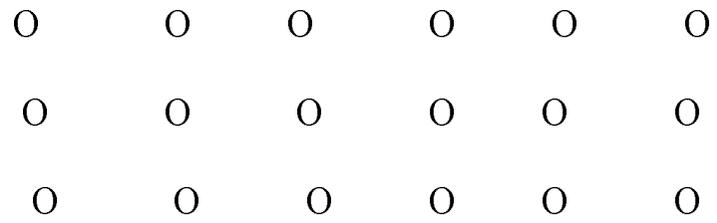
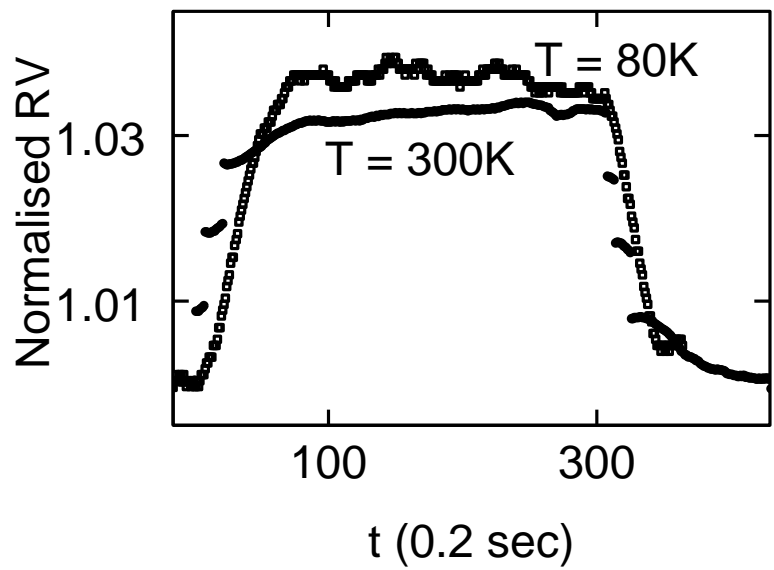
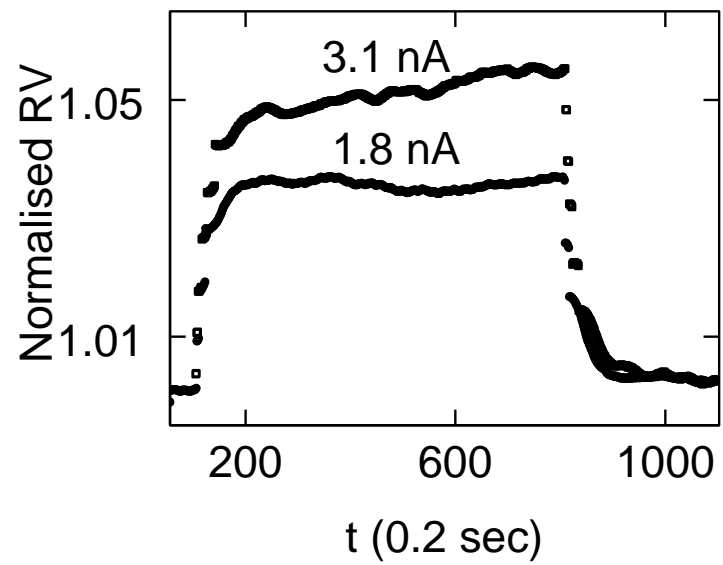
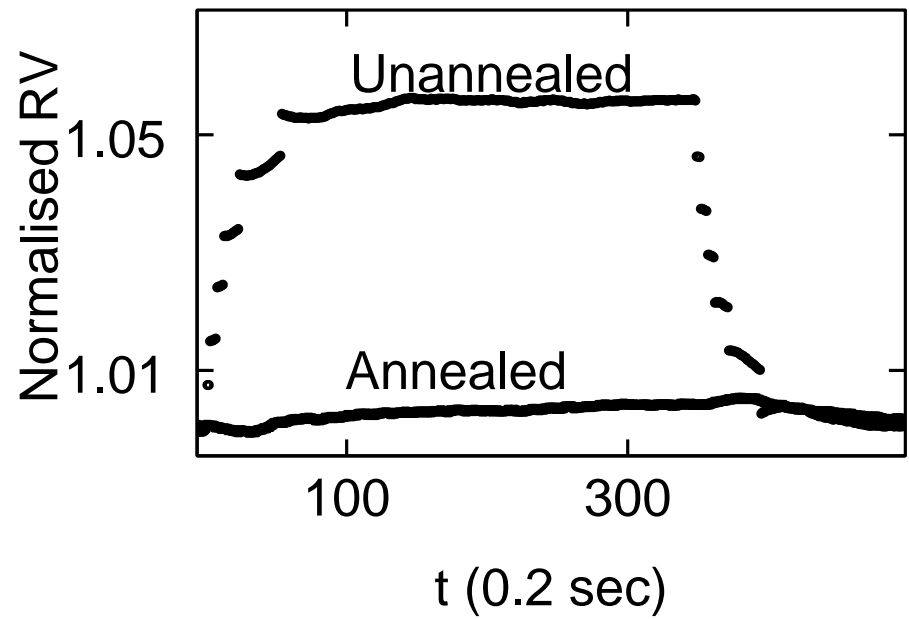
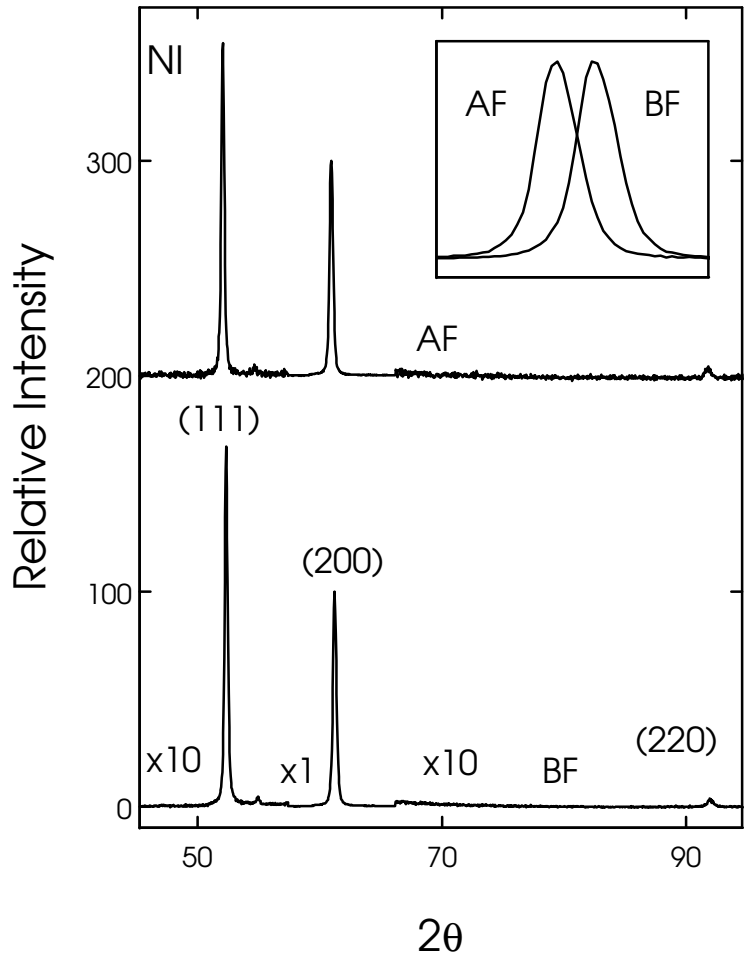
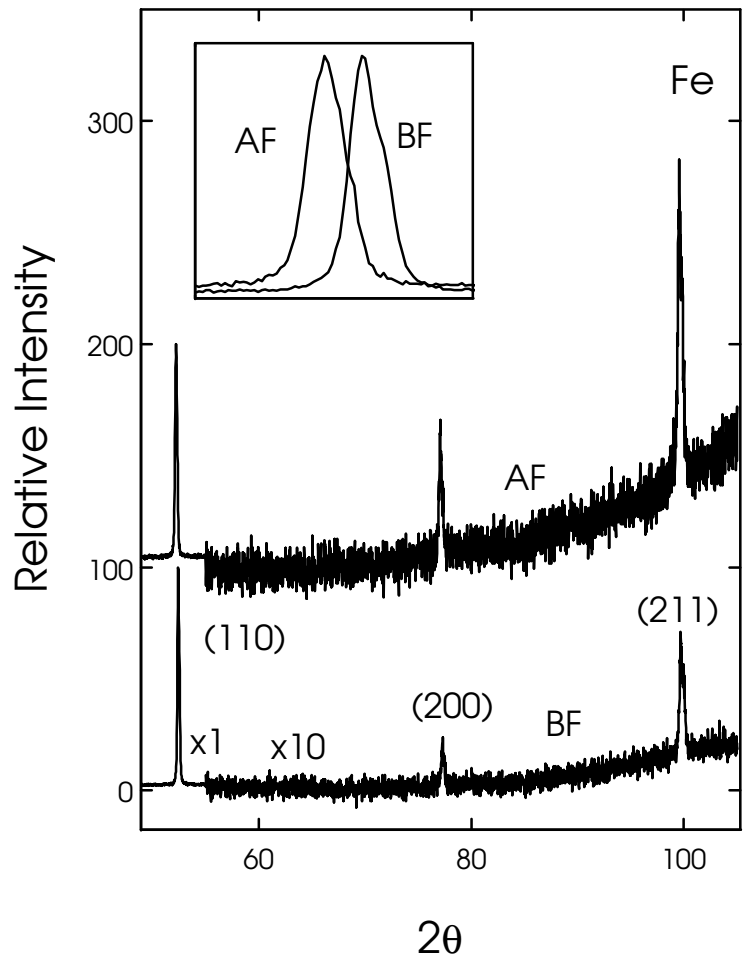
\end{document}